\documentclass[aps,prb,superscriptaddress,twocolumn]{revtex4-1}

\usepackage{graphics}
\usepackage{graphicx} 
\usepackage{dcolumn}
\usepackage{bm}
\usepackage{epstopdf}
\usepackage[usenames, dvipsnames]{color}

\begin{document}


\title{Interplay of spin-orbit coupling and hybridization in Ca$_3$LiOsO$_6$ and Ca$_3$LiRuO$_6$}



\author{S.~Calder}
\email{caldersa@ornl.gov}
\affiliation{Quantum Condensed Matter Division, Oak Ridge National Laboratory, Oak Ridge, Tennessee 37831, USA.}

\author{D.~J.~Singh}
\affiliation{Department of Physics and Astronomy, University of Missouri,Columbia, MO 65211-7010, USA}

\author{V.~O.~Garlea}
\affiliation{Quantum Condensed Matter Division, Oak Ridge National Laboratory, Oak Ridge, Tennessee 37831, USA.}

\author{M.~D.~Lumsden}
\affiliation{Quantum Condensed Matter Division, Oak Ridge National Laboratory, Oak Ridge, Tennessee 37831, USA.}

\author{Y.~G.~Shi}
\affiliation{Beijing National Laboratory for Condensed Matter Physics and Institute of Physics, Chinese Academy of Sciences, Beijing 100190, China}

\author{K.~Yamaura}
\affiliation{Research Center for Functional Materials, National Institute for Materials Science, 1-1 Namiki, Tsukuba, Ibaraki 305-0044, Japan}
\affiliation{Graduate School of Chemical Sciences and Engineering, Hokkaido University, North 10 West 8, Kita-ku, Sapporo, Hokkaido 060-0810, Japan}

\author{A.~D.~Christianson}
\affiliation{Quantum Condensed Matter Division, Oak Ridge National Laboratory, Oak Ridge, Tennessee 37831, USA.}

 

\begin{abstract}
The electronic ground state of Ca$_3$LiOsO$_6$ was recently considered within an intermediate coupling regime that revealed $J$=3/2 spin-orbit entangled magnetic moments. Through inelastic neutron scattering and density functional theory we investigate the magnetic interactions and probe how the magnetism is influenced by the change in hierarchy of interactions as we move from Ca$_3$LiOsO$_6$ (5$d$$^3$) to Ca$_3$LiRuO$_6$ (4$d$$^3$). An alteration of the spin-gap and ordered local moment is observed, however the magnetic structure, N\'{e}el temperature and exchange interactions are unaltered. To explain this behavior it is necessary to include both spin-orbit coupling and hybridization, indicating the importance of an intermediate coupling approach when describing 5$d$ oxides.
\end{abstract}


\maketitle

\section{\label{sec:Introduction}Introduction}

Investigations of 4$d$ and 5$d$-based compounds in recent years have revealed an array of novel phenomena. For example non-trivial topological insulating states, Kitaev and Majorana Fermion realizations, proximate superconductivity, anomalously high magnetic ordering temperatures and novel insulating ground states \cite{PhysRevLett.102.017205, annurev-conmatphys-020911-125138}. Much of the focus has centered on iridates with 5$d^5$ electronic occupancy in which the presence of magnetism and Mott-like insulating behavior is considered in either the LS-coupling limit, employed typically for 3$d$ elements with weak spin-orbit coupling (SOC) or the $jj$-coupling limit, typically utilized for heavy 4$f$ or 5$f$ ions with strong SOC \cite{Khomskii_book}. The near-cubic crystal field and SOC in 5$d^5$ iridates are considered as breaking the $d$-manifold degeneracy to create a $J_{\rm eff}$=1/2 doublet ground state \cite{KimScience}, with Coulomb interactions and hybridization (i.e orbital overlap) entering as perturbations \cite{annurev-conmatphys-020911-125138}. This approach has proven effective in explaining much of the phenomena, however, general extension to other electronic occupancies has proven problematic. In particular 5$d^3$ oxides  would be expected to have quenched orbitals in this approach with SOC only entering as a third order perturbation. However,  increasing experimental evidence indicates a strong orbital contribution in their magnetic behavior \cite{PhysRevB.84.174431, PhysRevB.91.075133, calder2015_cd227, TaylorSpinGap}. 

To provide a more applicable model the 5$d$ electronic ground state was recently described within an intermediate coupling (IC) model that incorporates SOC, Coulomb interactions and hybridization on an equal footing rather than invoking the LS or $jj$ limits \cite{Alice_RIXS}. This approach yields the eigenvectors and eigenvalues of the electronic ground state and excited states by using Racah parameters to incorporate electron-electron repulsion into the model and allows for the extraction of the Hund's coupling energy ($J_h$), SOC ($\zeta$$\rm _{SOC}$) values and gives insights into the degree to which the compound deviates from pure ionic towards covalency, the so-called nephelauxetic effect \cite{Eisenstein, Racah_rev}. Specifically, a thorough modeling of the experimentally measured $d$-manifold splitting in 5$d^3$ Ca$_3$LiOsO$_6$ showed it to consist of a $J$=3/2 spin-orbit entangled ground state \cite{Alice_RIXS}. 

\begin{figure}[tb]
	\centering         
	\includegraphics[trim=0.5cm 2.8cm 0.5cm 0.0cm,clip=true, width=1.0\columnwidth]{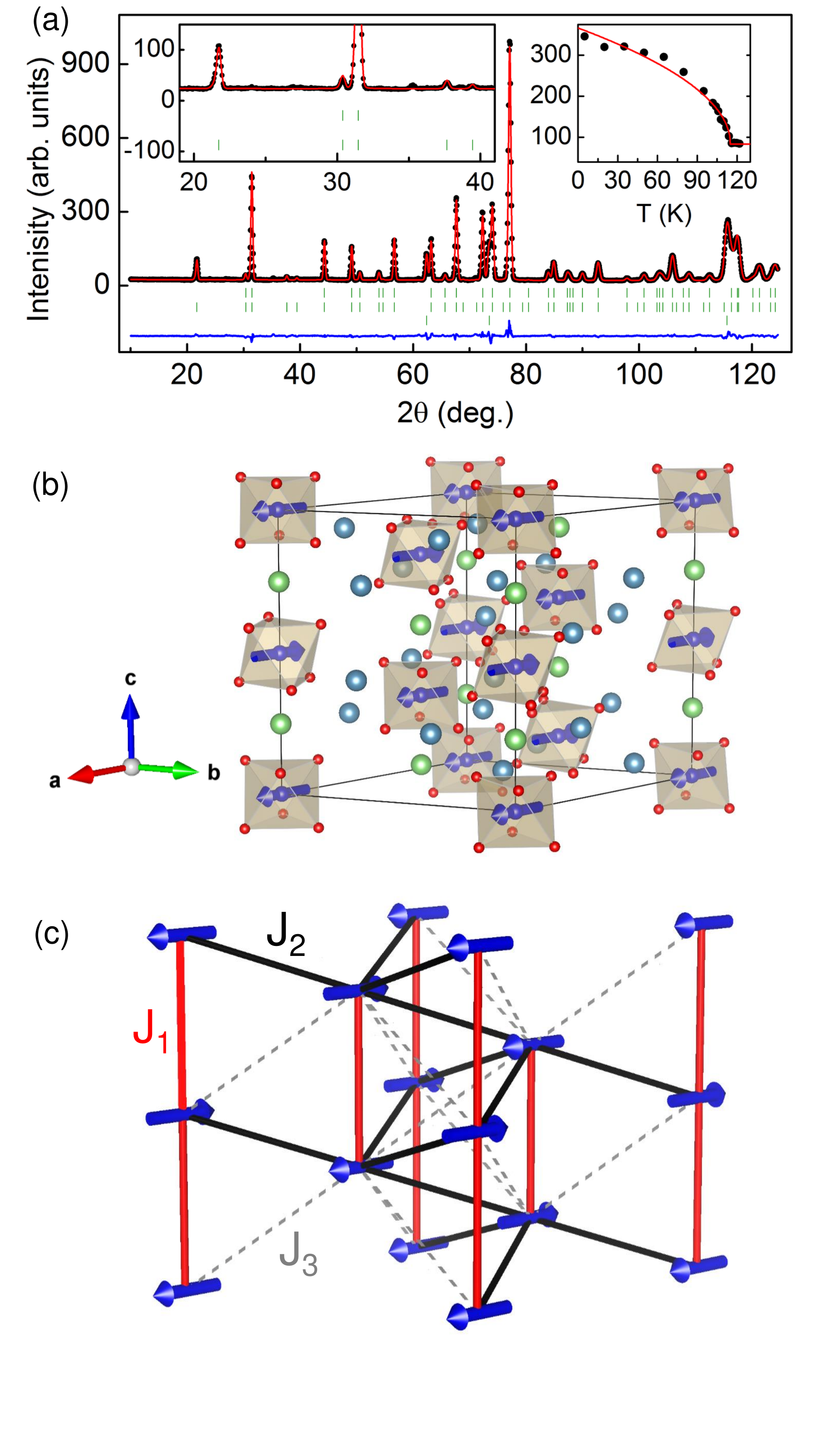}             
	\caption{\label{FigMagStr}(a) Neutron powder diffraction data (black) and model (red) for Ca$_3$LiRuO$_6$ at 4 K. Tick marks correspond to, from top to bottom, the crystal structural, magnetic structure and Al sample holder.  Inset: intensity of magnetic reflection at $2\theta$$=$22$^\circ$. (b) Ca$_3$LiOsO$_6$ and Ca$_3$LiRuO$_6$ form the hexagonal 2$H$-perovskite structure. Magnetic Os$^{5+}$/Ru$^{5+}$ ions (blue) are surrounded by six oxygen anions (red spheres). These octahedra are separated along the $c$-axis by Li$^+$ cations. Ca$^{2+}$ ions (grey spheres) provide the charge balance. (c) The magnetic structure and exchange interaction pathways are shown.}
\end{figure}

Ca$_3$LiOsO$_6$ is well suited as a canonical model system to investigate the IC regime in a crystalline inorganic compound due to the well isolated 5$d$ octahedra in the $2H$-perovskite structure, space group $R\bar3c$, $\#$167 shown in Fig.~\ref{FigMagStr}. The magnetism of Ca$_3$LiOsO$_6$ on this crystal structure is a rare example comprised of solely extended superexchange interactions (Os-O-O-Os) \cite{ShiJACS,KanInorChem}. This has an advantage in probing the IC model since it places hybridization in a more pronounced role compared to the greater studied 5$d$ oxides with superexchange interactions (Os-O-Os). Moreover, the single-ion ground state can be more robustly experimentally accessed since closer-range interactions that can potentially mask the ground state behavior in spectroscopic measurements are not present \cite{ament2011, Alice_RIXS}. While Ca$_3$LiOsO$_6$ is a good model system, the observed departures from the LS and $jj$ coupling limits in several 5$d$ compounds independent of whether they have either superexchange or extended superexchange interactions suggests a wide applicability of the IC model in 5$d$-based compounds as a way of gaining new physical insights. 

Here we investigate the magnetic exchange interactions between the $J$=3/2 moments in Ca$_3$LiOsO$_6$ with inelastic neutron scattering and contrast them with the 4$d$ analogue Ca$_3$LiRuO$_6$. The results are modeled within linear spin-wave theory to reveal dominant extended superexchange interactions. Despite the indistinguishable magnetic structure and ordering temperature of Ca$_3$Li$T$O$_6$ ($T$=Os,Ru) the emergence of an observable spin-gap in the excitation spectra going from 4$d^3$ to 5$d^3$ signifies the change of SOC and an alteration of behavior from the expected S=3/2 ground state in Ca$_3$LiRuO$_6$ to the $J$=3/2 electronic ground state in Ca$_3$LiOsO$_6$. By performing detailed density functional theory (DFT) we are able to explain the observed behavior when considering the combined and altering influence of SOC and hybridization. Collectively our results reveal the importance of an IC approach in describing the magnetism of 5$d^3$ compounds and their novel emergent behavior.

\section{\label{sec:Expt_Theory}Experimental and theoretical details}

\subsection{Sample preparation}

Polycrystalline Ca$_3$LiOsO$_6$ and Ca$_3$LiRuO$_6$ were prepared using solid state techniques as described in Ref.~\onlinecite{ShiJACS}. 5g of Ca$_3$Li$T$O$_6$ were loaded in identical Al cylindrical holders for the neutron scattering measurements.

\subsection{Neutron powder diffraction}

Neutron powder diffraction measurements were performed on the HB-2A powder diffractometer at the High Flux Isotope Reactor (HFIR), Oak Ridge National Laboratory (ORNL) \cite{Garlea2010}. Measurements on Ca$_3$LiRuO$_6$ were performed with a wavelength of 2.41~\AA~ at 4 K, 90 K, 125 K, 140 K and 250 K to follow the development of magnetic ordering. The magnetic structure was modeled using Fullprof \cite{Fullprof}  through a representational analysis using SARAh \cite{sarahwills}, with the results compared against magnetic symmetry using the Bilbao crystallographic server \cite{Bilbao_Mag}. 

\subsection{Inelastic neutron scattering}

Inelastic neutron scattering measurements were performed on polycrystalline samples on the ARCS and HYSPEC time of flight spectrometers at the Spallation Neutron Source (SNS), ORNL.  Incident energies of 20, 50, 80, 120 meV were used on ARCS to cover the full magnetic excitation energy range at several temperatures through $\rm T_N$. A lower incident energy of $E_i$=13 meV was used on HYSPEC for improved resolution of 0.3 meV.

\subsection{Density functional theory calculations}

Density functional theory (DFT) calculations were performed using the generalized gradient approximation (GGA) of Perdew, Burke and Ernzerhof (PBE) and the general potential linearized augmented planewave (LAPW) method \cite{singh-book} as implemented in the WIEN2k code \cite{wien2k}. We used LAPW sphere radii of 1.55 bohr for O, 2.0 bohr for Li and 2.1 bohr for Ca, Ru and Os. We used the standard LAPW basis set plus local orbitals for the semicore states, including the semicore $s$ and $p$ states of both Ru and Os. For the structure we used the experimentally determined lattice parameters and relaxed the internal atomic positions using the PBE GGA \cite{pbe}. We used this structure for calculating electronic and magnetic properties as discussed below. All calculations included SOC, except for the structure relaxation, which was done including magnetism with the core states treated fully relativistically and the valence states treated in a scalar relativistic approximation. All the magnetic calculations, including ferromagnetic and various antiferromagnetic states, yield similar moments equivalent to half filled high spin $t_{2g}$ orbitals, $S$=3/2, with a reduction in the spin moments due to SOC, and orbital moment opposite to the spin-moment following Hund's rules.

\section{\label{sec:Res_Discuss}Results and Discussion}


\begin{figure}[tb]
	\centering                      
	\includegraphics[trim=2.5cm 4.5cm 2.0cm 0.8cm,clip=true, width=1.0\columnwidth]{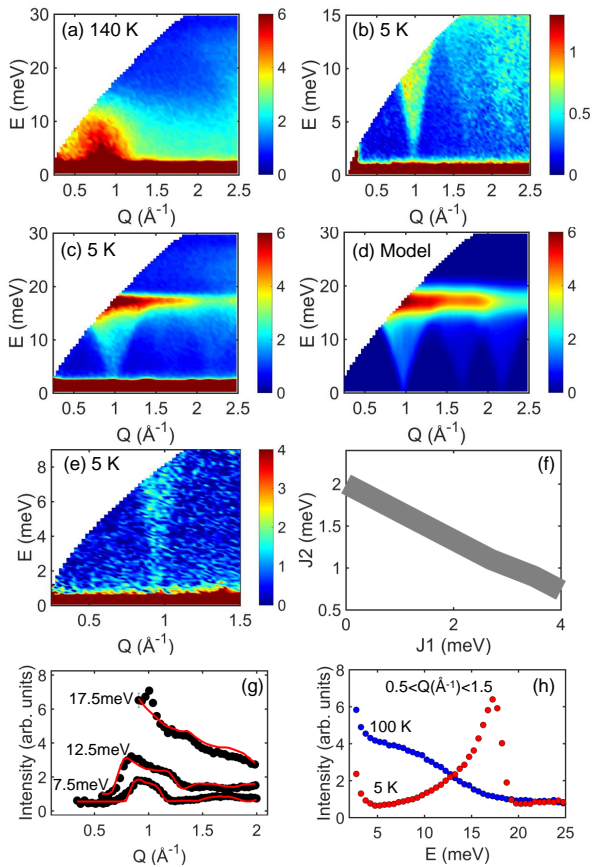}
	\caption{\label{FigRuARCS} Inelastic neutron scattering measurements on Ca$_3$LiRuO$_6$ at temperatures and incident energies of (a) $T$=140 K, $E_i$=50 meV, (b) $T$=5 K, $E_i$=20 meV, (c) $T$=5 K, $E_i$=50 meV, (d) Calculated spectra, (e) $T$=5 K, $E_i$=13 meV. (f) Antiferromagnetic $J_1$ and $J_2$ values that correspond to energy of the top of the band. (g) Measured (black circles) and calculated (red line) constant-E cuts. (h) Constant Q-cuts.}
\end{figure}

\begin{figure}[tb]
	\centering                   
	\includegraphics[trim=3.2cm 1.5cm 3.5cm 0.7cm,clip=true, width=1.0\columnwidth]{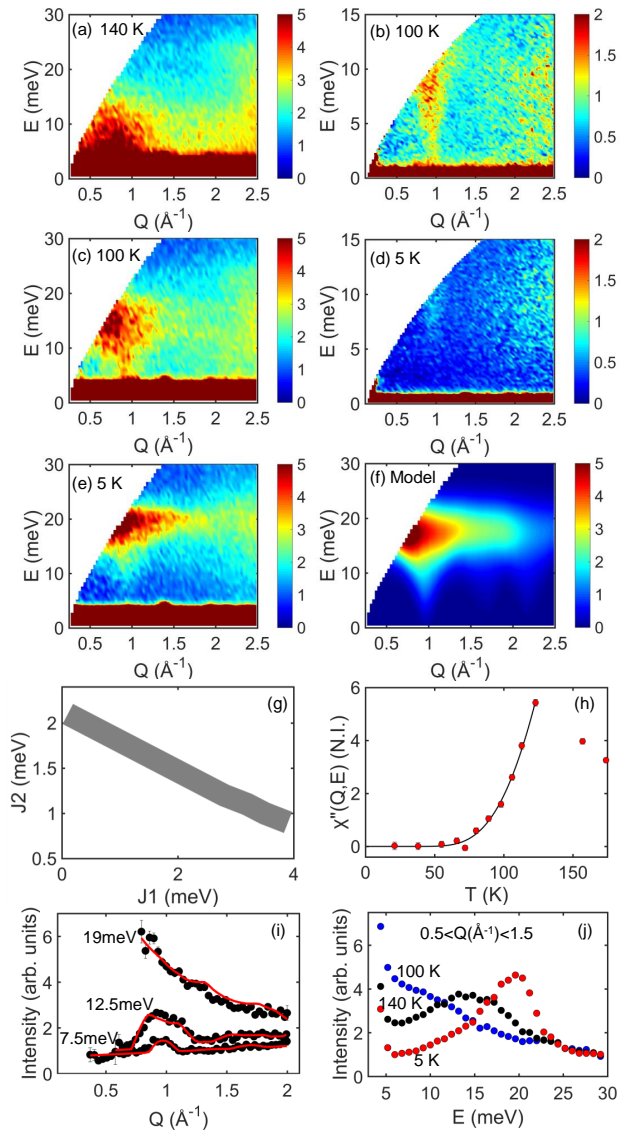}
	\caption{\label{FigOsARCS} Inelastic neutron scattering measurements on Ca$_3$LiOsO$_6$ at temperatures and incident energies of (a) $T$=140 K, $E_i$=80 meV, (b) $T$=100 K, $E_i$=20 meV, (c) $T$=100 K, $E_i$=80 meV, (d) T=5 K, $E_i$=20 meV and (e) $T$=5 K and $E_i$=80 meV. (f) Calculated spectra. (g) Antiferromagnetic $J_1$ and $J_2$ values that correspond to energy of the top of the band. (h) Bose corrected intensity in region 4-7 meV and Q=0.9-1.1$\rm \AA^{-1}$ to follow evolution of spin gap. (i) Measured (black circles) and calculated (red line) constant-E cuts. (j) Constant Q-cuts.}
\end{figure}

Previous characterizations of Ca$_3$LiOsO$_6$ and Ca$_3$LiRuO$_6$ have shown that both compounds should undergo long range magnetic order around the same temperature and share indistinguishable values of $\theta \rm _{CW}$ and $\mu \rm _{eff}$ from Curie-Weiss fits to the magnetic susceptibility   \cite{DARRIET1997139, ShiJACS}. While the long range magnetic structure and ordering temperature of Ca$_3$LiOsO$_6$ has been probed \cite{PhysRevB.86.054403}, no such measurements exist for Ca$_3$LiRuO$_6$. We therefore performed neutron powder diffraction measurements on Ca$_3$LiRuO$_6$ on the HB-2A diffractometer at HFIR in Fig.~\ref{FigMagStr}(a). An identical set of magnetic reflections are observed for Ca$_3$LiRuO$_6$ at low temperature (4 K) as found for Ca$_3$LiOsO$_6$ and can be indexed to the same $k$=(0,0,0) propagation vector. We follow the procedure as described previously for Ca$_3$LiOsO$_6$ in Ref.~\onlinecite{PhysRevB.86.054403} to assign the magnetic structure. To further constrain the magnetic structure we note the similar behavior observed in the related 2$H$-compounds from single crystal measurements in Ref.~\onlinecite{PhysRevB.94.134404}. Therefore, we assign the magnetic symmetry as given by $C2'/c'$ ($\#$15.89). The spins are found to reside in the $ab$-plane, although within the limits of our powder samples we cannot constrain the direction to a unique axis. The intensity of the magnetic peaks were larger for Ca$_3$LiRuO$_6$ compared to Ca$_3$LiOsO$_6$ and this allowed any $c$-axis component, allowable from a magnetic symmetry analysis \cite{Gallego:ks5532}, to be probed with more robustly. However, as for Ca$_3$LiOsO$_6$ no $c$-axis component was observable. The ordered magnetic moment is 2.8(1)$\mu\rm _B$/Ru ion, close to the full ordered moment of 3$\mu\rm _B$ and notably larger than the 2.2(1)$\mu\rm _B$/Os ion \cite{PhysRevB.86.054403}. Fitting the intensity of a magnetic reflection as a function of temperature to a power law yielded $\rm T_N$=117.0(8) K, see Fig.~\ref{FigMagStr}(a), giving an identical ordering temperature for Ca$_3$LiRuO$_6$  and Ca$_3$LiOsO$_6$ \cite{PhysRevB.86.054403}. During the measurements we tested the suggestion that an anomaly above the transition temperature may be indicating a wide temperature region of short range correlations \cite{0953-8984-28-23-236001}. However, no such correlations were observed.

Having established that both Ca$_3$LiOsO$_6$ and Ca$_3$LiRuO$_6$ share the same magnetic structure and ordering temperature, the highest such ordering temperature for the $2H$-perovskite structure, we now probe the collective magnetic excitations with inelastic neutron scattering.  The measurements on Ca$_3$LiRuO$_6$ and Ca$_3$LiOsO$_6$ are shown in Figs.~\ref{FigRuARCS}-\ref{FigOsARCS} and reveal well defined excitations that follow temperature and Q dependence consistent with magnetic excitations. Inspecting the spectra immediately reveals key distinctions between the 4$d^3$ and 5$d^3$ analogues. The most pronounced and significant difference in terms of the underlying physics is the apparent presence of a gapless spin excitation in Ca$_3$LiRuO$_6$ but a gapped excitation for Ca$_3$LiOsO$_6$ at the magnetic zone center. In Fig.~\ref{FigOsARCS}(h) we follow the Bose corrected intensity to reveal the development of a spin gap in going through $\rm T_N$ by fitting the $T$$<$$T_N$ results to $\chi''(T) \propto(-\Delta/k_BT)$. Conversely for Ca$_3$LiRuO$_6$ even with the best resolution of 0.3 meV, Fig.~\ref{FigRuARCS}(e), there is no observable spin-gap. We will return to the impact of this distinction below. A further contrast in the scattering is the overall bandwidth appears similar, but the top of the band energy is higher for Ca$_3$LiOsO$_6$ compared to Ca$_3$LiRuO$_6$. In Fig.~\ref{FigRuARCS} and \ref{FigOsARCS} we performed constant Q-cuts to define the top of the band. We find values of E=19.5(7) meV for Ca$_3$LiRuO$_6$ and  E=17.3(5) meV for Ca$_3$LiOsO$_6$. The final contrast between the spectra for Ca$_3$LiRuO$_6$ and Ca$_3$LiOsO$_6$, the difference in the relative broadness of the S(Q,$\omega$) scattering, can be assigned to the more delocalized nature of the 5$d$ magnetism compared to 4$d$.


To provide a quantitative description of the excitation spectra we invoke linear spin wave theory and use the Hamiltonian $\mathcal{H}=\sum_{i,j}J_{ij} \mathbf{S}_i\cdot\mathbf{S}_j + \mathcal{H}_A$, where $J_{ij}$ are the Heisenberg exchange  interactions and $\mathcal{H}_A$ encompasses the anisotropic terms from single-ion anisotropy, anisotropic exchange and Dzyaloshinskii-Moriya (DM) interaction, allowable from the broken inversion symmetry. To initially guide our results we began with parameters previously calculated for Ca$_3$LiOsO$_6$ \cite{KanInorChem}.  Ref.~\onlinecite{KanInorChem} assigns interactions out to next-next nearest neighbor, $J_3$, with the hierarchy $J_1$ $>$ $J_2$ $\gg$ $J_3$. Additionally we performed further detailed DFT calculations to aid the understanding of our experimental results, discussed in more detail below. Our DFT calculations indicate that while further neighbor interactions ($J_3$) may be present, a picture of nearest neighbor antiferromagnetic interactions, $J_1$ and $J_2$,  captures the essential physics. This is consistent with Ref.~\onlinecite{KanInorChem} that notes that only $J_1$ and $J_2$ are required to explain the three-dimensional antiferromagnetic ordering and moreover explain the high ordering temperature, T$\rm _N$=117 K, of these compounds.  Therefore to allow for a tractable solution we only include $J_1$ and $J_2$ interactions, as represented in Fig.~\ref{FigMagStr}, with the hierarchy determined in Ref.~\onlinecite{KanInorChem} of $J_1$ $>$ $J_2$, as a minimal model for three-dimensional magnetic ordering. To account for the spin-gap in the spectra of  Ca$_3$LiOsO$_6$ we incorporated $\mathcal{H}_A$. We utilized a single-ion term $\sum_{i,\alpha} -D_{\alpha} (S_i^{\alpha})^2$ with $\alpha$ fixed along the spin direction in the plane, but note that using single-ion, as opposed to DM or anisotropic exchange, is arbitrary and assigning the relative contributions from all three is beyond the limits of the data. To determine the $J_1$ and $J_2$ exchange interactions we took constant-E and constant-Q cuts of the data and then fit these using a least squares analysis of the calculated powder averaged model of the spin wave dispersion using spinW \cite{toth2015}. A constant scale factor was included in the model to simulate the intensity of the scattering and this remained fixed for all presented calculations for each compound. Throughout we use $S$=3/2 as obtained from magnetic susceptibility, rather than reduced ordered local moments measured with neutron diffraction, and report $J$ exchange values rather than $SJ$. To account for the broader scattering in Ca$_3$LiOsO$_6$ relative to Ca$_3$LiRuO$_6$ we incorporate an artificial constant broadening term in energy. For Ca$_3$LiRuO$_6$ there is no spin-gap and therefore no experimental evidence for SOC creating anisotropic magnetic spins. Conversely for Ca$_3$LiOsO$_6$ a spin-gap is observed therefore we add this term to the Hamiltonian through an anisotropic single-ion term of 0.15 meV.  

For both compounds there is a range of exchange interactions that match both the top of the band energy and reproduce the experimental spectra adequately, shown in Fig.~\ref{FigRuARCS}(f) and Fig.~\ref{FigOsARCS}(g) for Ca$_3$LiRuO$_6$ and Ca$_3$LiOsO$_6$, respectively. However we note that solutions for $J_1$ $<$ $J_2$ are inconsistent with theoretical predications and we reject these values \cite{KanInorChem}. We found exchange interactions of $J_1$=2.1(5) meV and $J_2$=1.1(5) meV for both Ca$_3$LiOsO$_6$ and Ca$_3$LiRuO$_6$ as producing the best fit. As one test of how reasonable these exchange values are we calculate the Curie-Weiss temperature $\theta$ given by mean-field theory \cite{KanInorChem}: $\theta$=[$S(S+1)/3k{\rm_B}] \sum_{i}z_i Ji \approx 5(2J_1+6J_2+6J_3)/4k{\rm_B}$. The exchange interactions yield $\theta=-157$ K. This value appears reasonable given the measured $\rm T_N=117$ K and the extracted $\theta_{CW}=260$ K from susceptibility. 

The analogous magnetic exchange interactions naturally explain the indistinguishable magnetic structure and ordering temperature of Ca$_3$LiOsO$_6$ and Ca$_3$LiRuO$_6$. However, the question arises: why does this occur despite the observable difference in SOC, that controls magnetic-anisotropy, and the ordered magnetic moment sizes. To resolve this question and access the underlying phenomena we discuss DFT calculations performed for various magnetic configurations. The lowest energy magnetic ordering was nearest neighbor antiferromagnetism and the magnetocrystalline anisotropy gives an easy plane perpendicular to the rhombohedral axis for both compounds, in accord with experiment for both Ca$_3$LiOsO$_6$ \cite{PhysRevB.86.054403} and Ca$_3$LiRuO$_6$, Fig.~\ref{FigMagStr}. For the ground state antiferromagnetic order, the energy to switch the magnetization from the easy plane to the hard axis is 2.3 meV/Os for Ca$_3$LiOsO$_6$ but much reduced to 0.2 meV/Ru for Ca$_3$LiRuO$_6$. The anisotropy in the easy plane is below the precision of the calculations ($<$ 0.01 meV/Os). The dominance of hard-axis anisotropy indicates that the $\mathcal{H}_A$ term likely has strong contributions from DM and anisotropic exchange, however the specific nature and combination is beyond the limit of the results presented.

\begin{figure}[tb]
	\centering                      
	\includegraphics[width=\columnwidth]{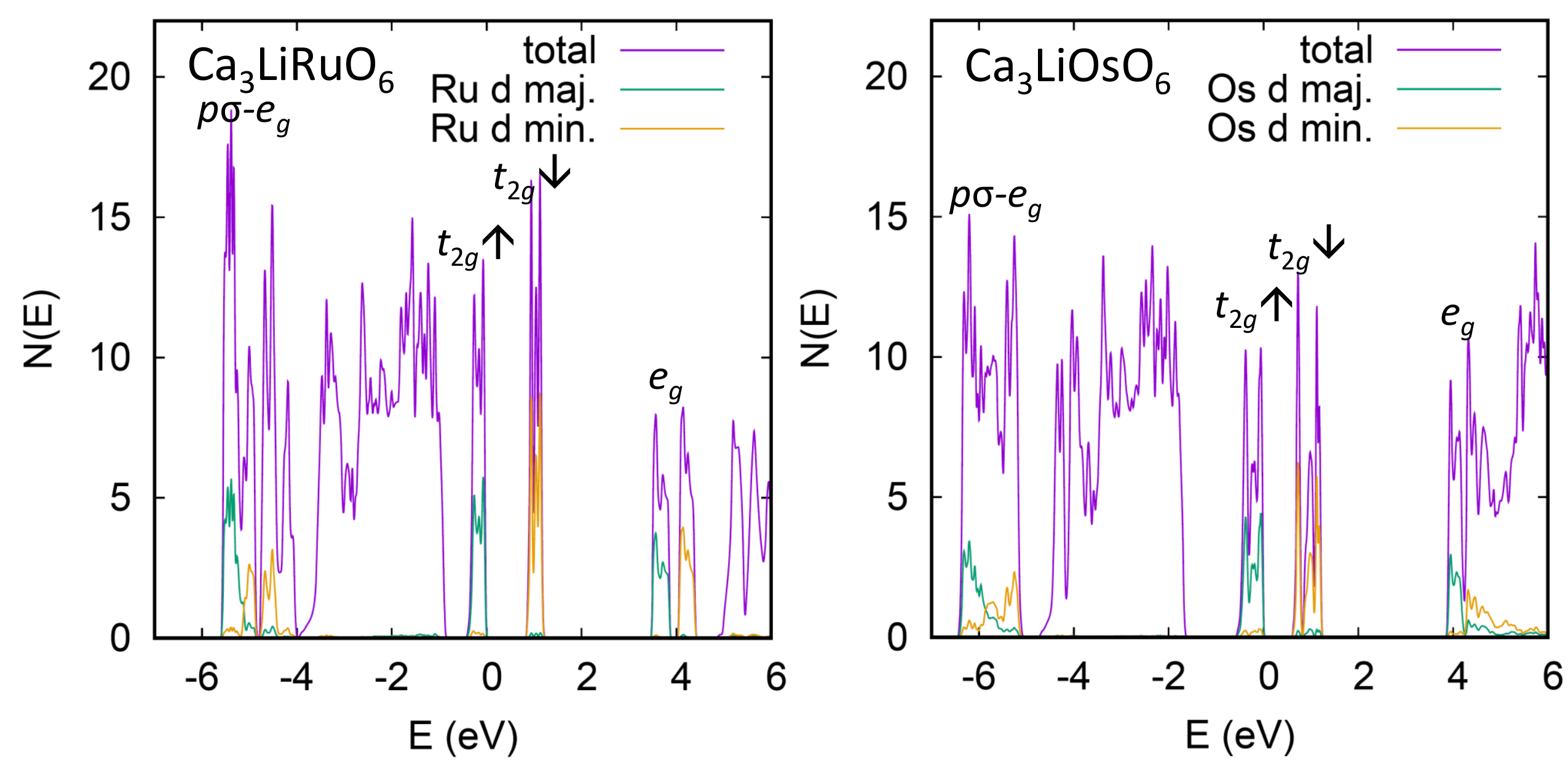}
	\caption{\label{FigcalcINS_DOS} Density of states per formula unit for Ca$_3$LiRuO$_6$ (left) and Ca$_3$LiOsO$_6$ (right) including SOC with projections of majority (maj.) and minority (min.) metal $d$ character onto the LAPW spheres. The main crystal field related peaks are indicated.}
\end{figure}

The calculated electronic densities of states (DOS) for the lowest energy magnetic structure is shown in Fig.~\ref{FigcalcINS_DOS}. The $d$ orbitals give rise to clearly separated crystal field and exchange split levels. On a qualitative level the DOS for Ca$_3$LiOsO$_6$ is similar to the $J$=3/2 ground state observed observed experimentally with the two main $t_{2g}$ peaks in the DOS separated by roughly the same energy as the two main peaks measured by RIXS \cite{Alice_RIXS}, with each of these peaks being further split into two visible subpeaks. The $t_{2g}$ exchange splittings of 1.2 eV are similar in the two compounds. Interestingly, the exchange splitting of the $e_g$ states is significantly lower, $\sim$0.6 eV. This is another indicator of the importance of $d$-$p$ hybridization, where the stronger covalency of the $e_g$ states works against the Hund's rule exchange coupling that would favor equal exchange splittings for all $d$ orbitals.

The crystal field in oxides comes primarily from hybridization between metal $d$ orbitals and ligand $p$ orbitals. The lower lying $t_{2g}$ in an octahedral crystal field are from $\pi$ antibonding states, while the $\sigma$ antibonding $e_g$ level lies at higher energy due to the stronger $\sigma$ hybridization. Very large crystal field splittings of $\sim$3.6 eV and $\sim$4.0 eV for $T$=Ru and Os, respectively, are obtained, that for Ca$_3$LiOsO$_6$ agrees with RIXS measurements \cite{Alice_RIXS}. This indicates very strong hybridization of the $d$ orbitals and the O $p$ states. Additionally, the DOS shows substantial non-$d$ character in the $d$ bands as well as strong $d$ character at the bottom of the O $p$ valence bands. This is from $p$$\sigma$-$e_g$ bonding states. This mixture of metal $d$ and O $p$ states is one reason for obtaining reduced magnetic moments in the neutron measurements. 

Thus in both compounds a sizable portion of the moment is distributed among the O ions, which are polarized in the same direction as the $T$ ions that they surround. This is a consequence of the considerable hybridization between 4$d$ or 5$d$ orbitals and the $p$ orbitals of the surrounding O. The sizable polarization of the O ions means that large interactions can be expected along the O-O bridges connecting the octahedra. Therefore it is instructive to view the magnetic structure and interactions as based on ($T$O$_6$)$^{7-}$ units, stabilized by the Ewald field of the cations, and interacting through their contacts. This is similar to the picture developed for some ruthenate and osmate double perovskites \cite{mazin-ruthenate, TaylorSpinGap, morrow}. In Ca$_3$Li$T$O$_6$  each O in the $T$O$_6$ units has short distances to three O in different neighboring units. For $T$=Os (Ru), these short distances are  2.87 \AA (2.85 \AA), 2.94 \AA (2.94 \AA), and 3.10 \AA (3.10 \AA)  with the shorter two to octahedra in the basal plane and the long distance to the next octahedron in the $c$-axis direction. From this point of view, the structure does not show 1D Os chains often associated with this crystal structure. 

For ferromagnetic order, without SOC, the spin moments are exactly 3$\mu_B$ per formula unit, of which the Ru $d$ contribution, measured by the moment in the Ru LAPW sphere is only 1.73$\mu_B$. While Ru $d$ orbitals extend beyond the 2.1 bohr sphere radius, most of the missing moment is from O, as each of the six O spheres contain 0.10$\mu_B$, and the $p$ orbitals of O$^{2-}$ also are extended beyond the 1.55 bohr spheres. In the case of Os, there is 1.62$\mu_B$ inside the Os LAPW sphere. Single crystal neutron measurements would be of interest to probe for the O moment.  With SOC, for the ground state antiferromagnetic order, the Ru spin moment, measured in the same way is almost the same, 1.71$\mu_B$, while the Os moment is significantly reduced by SOC to 1.50$\mu_B$. The orbital moments are -0.12 and -0.02$\mu_B$ for Os and Ru, respectively, and opposite to the spin-moment following Hund's rule.

The energy difference between ferromagnetic and nearest neighbor antiferromagnetic ordering without SOC is 0.074 eV/Ru for Ca$_3$LiRuO$_6$ and 0.090 eV/Os for Ca$_3$LiOsO$_6$. Introducing SOC has little effect for Ca$_3$LiRuO$_6$ yielding a value of 0.073 eV/Ru. However, SOC has the effect of significantly reducing the energy difference to 0.079 eV/Os for Ca$_3$LiOsO$_6$. Thus SOC not only reduces the moment in Ca$_3$LiOsO$_6$, but it also reduces the ordering energy to correspond closely to Ca$_3$LiRuO$_6$. Therefore, the similar ordering energies including SOC for the two compounds can be assigned to explain the similar N\'{e}el temperatures and magnetic structure. This indicates that tuning SOC and hybridization, due to the  altered influence these interactions have between 4$d^3$ to 5$d^3$ ions, is a route to traverse from $S$=3/2 to $J$=3/2 magnetism. Direct measurements of Ca$_3$LiRuO$_6$ to access the electronic ground state would be of interest to probe this cross-over.

\section{\label{sec:Conclusion}Conclusion}

Ca$_3$LiOsO$_6$ provides a model compound to investigate the IC regime and in particular $J$=3/2 spin-orbit entangled moments in 5$d^3$ compounds, with Ca$_3$LiRuO$_6$ offering an analogue with an altered hierarchy of competing interactions. An understanding of the magnetic properties reported necessitates the inclusion of SOC and strong hybridization. The hybridization mediates the extended magnetic interactions and increases the ordering temperature while SOC has the effect of reducing the moment and ordering energy. This competition leads to a surprising cancellation of energetics when going from 4$d$ to 5$d$ and results in Ca$_3$LiOsO$_6$ and Ca$_3$LiRuO$_6$ showing analogous magnetic ordered structures and transition temperatures. This observation has implications beyond these compounds and into 4$d$ and 5$d$ materials in general where an understanding of the magnetic interactions is of great current interest. In particular, while the well isolated OsO$_6$ octahedra in Ca$_3$LiOsO$_6$ allows for more direct experimental access to the $J$=3/2 ground state several features of related 5$d^3$ compounds, such as reduced moments, high ordering temperature and strong magnetic anisotropy, suggest a broader applicability of an IC approach will result in new understandings of materials where the magnetic ions are less isolated but SOC and hybridization are large.

\begin{acknowledgments}
	This research used resources at the High Flux Isotope Reactor and Spallation Neutron Source, a DOE Office of Science User Facility operated by the Oak Ridge National Laboratory. The work was supported in part by the Japan Society for the Promotion of Science (JSPS) through Grants-in-Aid for Scientific Research (15K14133 and 16H04501) and JSPS Bilateral Open Partnership Joint Research Projects. Y.G.S acknowledges support from the National Natural Science Foundation of China (Grant Nos. 11474330, 11774399). This manuscript has been authored by UT-Battelle, LLC under Contract No. DE-AC05-00OR22725 with the U.S. Department of Energy. The United States Government retains and the publisher, by accepting the article for publication, acknowledges that the United States Government retains a non-exclusive, paidup, irrevocable, world-wide license to publish or reproduce the published form of this manuscript, or allow others to do so, for United States Government purposes. The Department of Energy will provide public access to these results of federally sponsored research in accordance with the DOE Public Access Plan(http://energy.gov/downloads/doepublic-access-plan).
\end{acknowledgments}


%

\end{document}